# Single Cu Atom Sites on $Co_3O_4$ Activate Interfacial Oxygen for Enhanced Reactivity and Selective Gas Sensing at Low Temperature

Hamin Shin[1], Matteo D'Andria[1], Jaehyun Ko[1], Dong-Ha Kim[2], Frank Krumeich[3], Andreas T. Güntner[1]*

[1]Human-centered Sensing Laboratory, Department of Mechanical and Process Engineering, ETH Zürich, CH-8092, Zürich, Switzerland

[2]Department of Materials Science and Chemical Engineering, Hanyang University, Ansan 15588, Republic of Korea

[3]Department of Chemistry and Applied Biosciences, Laboratory of Inorganic Chemistry, ETH Zürich, CH-9083 Zürich, Switzerland

*Corresponding author e-mail: andregue@ethz.ch

**Abstract**

Controlling the redox landscape of transition metal oxides is central to advancing their reactivity for heterogeneous catalysis or high-performance gas sensing. Here we report single Cu atom sites (1.42 wt%) anchored on $Co_3O_4$ nanoparticles ($Cu_1$-$Co_3O_4$) that dramatically enhance reactivity and molecular sensing properties of the support at low temperature. The $Cu_1$ are identified by X-ray adsorption near edge structure and feature strong metal–support interaction between $Cu^{2+}$ and $Co_3O_4$, as revealed by X-ray photoelectron spectroscopy. The ability of $Cu_1$ to form interfacial Cu–O–Co linkages strongly reduces the temperature of lattice oxygen activation compared to CuO nanoparticles on $Co_3O_4$ ($CuO_{NP}$-$Co_3O_4$), as demonstrated by temperature-programmed reduction and desorption analyses. To demonstrate immediate practical impact, we deploy such $Cu_1$-$Co_3O_4$ nanoparticles as chemoresistive sensor for formaldehyde vapor that yields more than an order of magnitude higher response than $CuO_{NP}$-$Co_3O_4$ and consistently outperforms state-of-the-art sensors. That way, formaldehyde is detected down to 5 parts-per-billion at 50% relative humidity and 75 °C with excellent selectivity over various critical interferents. These results establish a mechanistic platform for activating redox-active supports using single-atom isolates of non-noble nature that yield drastically enhanced and well-defined reactivity to promote low-temperature oxidation reactions and selective analyte sensing.

# 1. Introduction

Controlling the spatial distribution and size of active metal species on catalyst supports offers unprecedented opportunities for the tuning of surface reactivity[1-3] with immediate practical impact on heterogenous catalysis,[4] energy storage[5] or molecular sensing.[6] In particular, down-sizing from metal nanoparticles (NPs) to single-atom sites (SAs) on metal oxides,[7] graphene derivatives[8] or metal-organic frameworks[9] has emerged as a transformative strategy, enabling maximal atom efficiency,[10,11] altered electronic structures,[12,13] and enhanced catalytic precision.[14] These attributes are advantageous, for instance, in chemoresistive gas sensors, where SAs offer the opportunity for well-defined surface reactivity and to explore different reaction pathways to optimize sensitivity and selectivity[15] to an analyte. While noble metal SAs such as Pt,[16,17] Pd,[18,19] and Au,[20,21] have been widely explored, non-noble transition metal SAs remain less investigated despite their abundance, thus cost-effectiveness, and diverse redox chemistries.

Among them, Cu-based SAs stand out as promising alternatives with proven performance for selective oxidation reactions[22,23] and electrocatalysis, such as oxygen reduction reaction.[24,25] Cu SAs are stabilized on supports by strong metal–support interactions (SMSI)[22] or local coordination environments[26] that modulate their oxidation states, thereby facilitating charge transfer and catalytic activation. While these studies highlight the catalytic potential of Cu SAs, most investigations have focused primarily on reactivity trends without deeply probing how the Cu dispersion state perturbs the redox landscape of the support. Furthermore, the role of SA-induced interfacial chemistry has rarely been addressed beyond model oxide systems, limiting broader understanding of how atomic Cu influences host materials with active lattice oxygen.

In particular, little is known about how atomically dispersed Cu modulates the electronic structure and oxygen activation capacity of redox-active supports compared to its nanoparticulate counterpart. The structural and electronic effects arising from different Cu configurations should be explored[27] to reveal structure–function relationships. Addressing these gaps will be critical for understanding how non-noble metal SAs interact with support lattices and for rationalizing their broader functional implications across redox-driven applications.[28]

Here, we explore how non-noble metal SAs fundamentally alter the redox properties of a transition metal oxide support through interfacial electronic interaction with immediate practical implication on low-temperature heterogeneous catalysis and chemoresistive sensing of environmental pollutants. Using flame-synthesized $Co_3O_4$ as a platform, we compare atomically dispersed Cu ($Cu_1$-$Co_3O_4$) with surface-deposited CuO nanoparticles ($Cu_{NP}$-$Co_3O_4$) at identical Cu surface loading to decouple the structural and electronic consequences of atomic-scale versus aggregated Cu species. A combination of X-ray spectroscopy and temperature-programmed analyses is utilized to identify the activation of lattice oxygen by Cu SAs and modifications to the Co redox landscape. To demonstrate the functional implications of our findings, we employ them exemplarily for formaldehyde gas sensing. This study bridges a mechanistic gap in non-noble metal SA research, and underscores the potential of atomic-scale engineering in developing efficient catalysts and sustainable sensing materials.

## 2. Results

### 2.1. Stabilization of Cu species on $Co_3O_4$: Structural Speciation

Flame-spray pyrolysis, a combustion-aerosol technique,[29] is utilized for the fabrication of $Co_3O_4$ nanoparticles with abundant oxygen vacancies, as demonstrated for other metal-oxides,[30,31] which are known to be one of the primary stabilization sites for single-atom sites (SAs).[32,33] Cu SAs are decorated by wet impregnation on the surface of $Co_3O_4$ nanoparticles, where a surface loading of 1.42 wt% Cu SAs is achieved, as determined by inductively coupled plasma mass spectrometry. This surface loading surpasses many previously reported values for Cu SAs on oxide supports prepared by conventional wet impregnation and sol-gel methods, which often result in limited dispersion.[10] For example, Cu loading was only to 0.34 wt% on ZSM-5[34] and 0.86 wt% on $CeO_2$-$TiO_2$,[22] likely due to the lower density of anchoring sites in such systems.

A high-resolution transmission electron microscopy image (**Figure 1a**, i) of the pure $Co_3O_4$ reveals well-defined crystalline nanoparticles with faceted shape and measured lattice spacing of 0.47 nm (see inset), corresponding to the (111) planes of spinel $Co_3O_4$ (JCPDS No. 42-1467). After the introduction of Cu SAs ($Cu_1$-$Co_3O_4$, **Figure 1a**, ii), the overall morphology of the $Co_3O_4$ nanoparticles and their crystal lattice remain unchanged. This suggests the successful anchoring of Cu species on the $Co_3O_4$ support, forming a $Cu_1$-$Co_3O_4$ heterostructure. For comparison, we also examined CuO NP-loaded samples (CuO$_{NP}$-Co$_3$O$_4$, **Figure 1a**, iii), where particles with lattice spacing of 0.28 nm are observed, which can be attributed to the (110) plane of CuO (JCPDS No. 48-1548).

Elemental mapping (EDS) further corroborates the presence of Cu SAs sites anchored on the Co$_3$O$_4$ lattice. Compared to the pure Co$_3$O$_4$ sample (**Figure 1b**, i), the Cu$_1$-Co$_3$O$_4$ (**Figure 1b**, ii and Figure S1) displays a homogeneous dispersion of Cu (green). The corresponding EDS spectra (Figures S2) confirm the presence of Cu alongside Co. In contrast, CuO$_{NP}$-Co$_3$O$_4$ (**Figure 1b**, iii) features localized Cu-rich regions that indicates the presence of clusters or nanoparticles attached to the Co$_3$O$_4$, despite identical Cu surface loading, as also confirmed by EDS spectra (Figure S2 versus Figure S3)

Powder X-ray diffraction (XRD) patterns (**Figure 1c**) are collected to analyze the crystal structure and phase composition of the pristine and Cu-loaded Co$_3$O$_4$ samples. All three samples show diffraction peaks that are associated with the spinel Co$_3$O$_4$ phase (squares, JCPDS No. 42-1467), indicating that the crystal structure of the support remains preserved upon Cu addition. The diffraction peaks of Cu$_1$-Co$_3$O$_4$ (**Figure 1c**, ii) closely resemble those of the pristine Co$_3$O$_4$ (**Figure 1c**, i), suggesting that the Cu SAs do not alter the crystalline bulk structure. In contrast, the CuO$_{NP}$-Co$_3$O$_4$ sample (**Figure 1c**, iii) shows additional reflections (triangles) at 2θ = 35.5° and 39.0° that can be indexed to the ($\bar{1}$11) and (111) planes of monoclinic CuO (JCPDS No. 48-1548), respectively, indicating the presence of crystalline CuO nanoparticles anchored (**Figure 1b**, iii) on the Co$_3$O$_4$ support.

To quantitatively assess the structural evolution of Co$_3$O$_4$ upon Cu SA anchoring, the Co$_3$O$_4$ crystal sizes of the samples are estimated from **Figure 1c** using the Scherrer equation. As shown in Figure S4, pristine Co$_3$O$_4$ exhibits the largest average crystallite size of 25.3 nm, in fair agreement with literature.[35] In case of Cu$_1$-Co$_3$O$_4$, the crystallite size decreases slightly to 23.1 nm, suggesting that the Cu SAs act as defects, thus interfering crystal growth during annealing (see Experimental section) possibly by stabilizing high-energy surfaces or introducing local strain.[36] In the CuO$_{NP}$-Co$_3$O$_4$ sample, the crystallite size further decreases to 21.3 nm, implying that surface-anchored CuO NPs (**Figure 1b**, iii) limit the coalescence and growth of Co$_3$O$_4$ crystallites possibly by acting as physical barrier.[16,37] This trend is in line with prior observations on CuO$_x$ clusters on Co$_3$O$_4$[35] and TiO$_x$[38] or Y$_x$O$_y$[39] on ZnO.

*2.2. Atomic Dispersion and Local Coordination of Cu Sites in $Cu_1$-$Co_3O_4$*

To elucidate the oxidation state and local coordination environment of Cu SAs anchored on $Co_3O_4$, X-ray absorption near-edge structure (XANES) and extended X-ray absorption fine structure (EXAFS) spectroscopy are performed at the Cu K-edge, along with near-edge X-ray absorption fine structure (NEXAFS) spectroscopy at the Cu $L_3$-edge (red lines, **Figure 2**). For comparison, CuO (black dashed line), $Cu_2O$ (dotted) and Cu foil (solid) are measured as reference samples, with their specifics being reported in the Experimental section.

As shown in the XANES spectra (**Figure 2a**, magnified in Figure S5), the white line position of $Cu_1$–$Co_3O_4$ lies between those of $Cu_2O$ and CuO, but more closely resembles the edge of CuO. This suggests that the Cu species predominantly exist in the $Cu^{2+}$ oxidation state. Notably, no discernible shoulder or pre-edge feature associated with metallic $Cu^0$ is observed, excluding the presence of Cu clusters or nanoparticles[26,40]. These results indicate that Cu is uniformly oxidized and does not form metallic aggregates on the $Co_3O_4$ surface. It is also worth noting that CuO and $Cu_2O$ display a distinct white line at the Cu K-edge, arising from dipole-allowed 1s → 4p transitions, as well as weak pre-edge shoulders arising from 3d–4p hybridization in distorted coordination environments.[41] In contrast, such spectral features are not observed for $Cu_1$–$Co_3O_4$, suggesting a more centrosymmetric oxygen coordination geometry around Cu, limiting 3d–4p mixing and thereby reducing white line asymmetry.[42] Such a local structure is consistent with the presence of atomically dispersed $Cu^{2+}$ centers on the $Co_3O_4$ support.

The NEXAFS spectrum (**Figure 2b**) further confirms the oxidized state of Cu in $Cu_1$-$Co_3O_4$. In fact, a pronounced white line feature at around 931 eV is observed at the Cu $L_3$-edge, attributed to the dipole-allowed $2p_{3/2}$ → 3d electronic transitions in $Cu^{2+}$.[43] The intensity and energy position of this feature are closely aligned with those of the CuO reference spectrum, indicating that the Cu species in $Cu_1$-$Co_3O_4$ adopt a similar oxidation state. However, the white line in $Cu_1$-$Co_3O_4$ appears slightly sharper and more symmetric compared to that of bulk CuO, suggesting that the single-atom Cu isolates experience a more homogeneous and possibly more symmetric ligand field.[44] This distinction implies that the $Cu^{2+}$ centers in $Cu_1$-$Co_3O_4$ are not part of a bulk-like CuO phase but are rather atomically dispersed, in agreement with **Figure 1**, and coordinated to surface oxygen atoms in the $Co_3O_4$ matrix.[42]

To probe the local coordination structure, Fourier-transformed (FT)-EXAFS analysis is carried out (**Figure 2c**). The FT-EXAFS spectrum of $Cu_1$-$Co_3O_4$ (red line) shows a dominant peak at approximately 1.5 Å, corresponding to the Cu–O first coordination shell.[45] Importantly, no distinct peaks are observed beyond this region, especially not at ~2.2 Å as with Cu foil (black solid line), that would indicate Cu–Cu scattering.[26] This absence of Cu–Cu scattering provides further evidence that the Cu atoms are atomically dispersed and not aggregated as nanoparticles or clusters. In fact, CuO (black dashed line) and $Cu_2O$ (black dotted line) exhibit distinct second-shell features above 2 Å due to Cu–Cu coordination in their extended lattice structures.

To gain further structural insight, quantitative EXAFS curve fitting is performed in R-space using a two-shell model comprising Cu–O and Cu–M (M = Co) scattering paths, based on the $Co_3O_4$ crystal framework (Figure S6). The fitting results show that the experimental data were well reproduced with this model, further supporting the local coordination of $Cu^{2+}$ species within the oxide matrix. The fitted parameters are summarized in Table S1. The coordination number of Cu–O is found to be approximately 3.5, suggesting a planar geometry. The fitted bond distance for Cu–O is 2.08 Å, consistent with literature values for $Cu^{2+}$–O bonds in oxide environments.[45]

Collectively, these results strongly indicate that Cu species in $Cu_1$-$Co_3O_4$ are present as atomically dispersed $Cu^{2+}$ centers, coordinated exclusively to oxygen atoms, with no detectable formation of metallic or oxide-based Cu aggregates. This local coordination environment is illustrated in **Figure 2d** (alternative view in Figure S7), showing a $Cu^{2+}$ center stabilized by oxygen ligands at the $Co_3O_4$ (001) facet, consistent with the configuration suggested by the EXAFS fitting. The atomically dispersed $Cu^{2+}$ sites stabilized on the flame-made $Co_3O_4$ support are expected to serve as unique active centers for catalytic and gas-sensing applications, as explored below.

*2.3. Surface Redox Properties and Metal–Support Interaction*

To explore how the dispersion state of Cu modulates the surface reactivity and the degree of SMSI, $H_2$-temperature-programmed reduction ($H_2$-TPR), formaldehyde-temperature-programmed desorption (formaldehyde-TPD), and X-ray photoelectron spectroscopy (XPS) analyses are conducted (**Figure 3**).

The $H_2$-TPR results (**Figure 3a**) reveal differences in reduction behavior depending on the Cu configuration. Both pristine $Co_3O_4$ (green) and $CuO_{NP}$-$Co_3O_4$ (blue) display nearly identical two-step reduction profiles, corresponding to the sequential reduction of $Co^{3+}$ to $Co^{2+}$ (~275 °C) and $Co^{2+}$ to $Co^0$ (~360 °C).[46] In $CuO_{NP}$-$Co_3O_4$, an additional minor reduction peak near 210 °C can be attributed to the surface reduction of CuO nanoparticles. This assignment is consistent with the $H_2$-TPR profile of bulk CuO (purple), which exhibits a two-step reduction starting at 170 °C with peak at 210 °C that correspond to the $Cu^{2+} \rightarrow Cu^+$ and $Cu^+ \rightarrow Cu^0$ transitions, respectively.[47] However, the stronger similarity with pristine $Co_3O_4$ implies that the presence of CuO NPs hardly affects the reducibility of the $Co_3O_4$ matrix. In stark contrast, $Cu_1$-$Co_3O_4$ exhibits a single reduction band with onset at ~170 and peak at ~275 °C, significantly earlier than in the other $Co_3O_4$-based samples and more pronounced than CuO. The low-temperature position of this band suggests the sequential reduction of $Co^{\delta+}$ at lower temperatures due to the presence of atomically dispersed Cu. This is reminiscent of the temperature shift observed in $Co_3O_4$ loaded with amorphous or sub-nanometer $CuO_x$ clusters, where a distinct reduction peak appeared at 160 °C, but only at ≥3 wt% Cu loading.[35] We suggest that the incorporation of atomically dispersed Cu in the $Co_3O_4$ matrix increases the mobility of surface lattice oxygen by modulating the local electronic structure and promoting the formation of oxygen vacancies, thereby facilitating redox processes at lower temperatures.

To evaluate how these structural changes influence surface adsorption properties, formaldehyde-TPD measurements are performed (**Figure 3b**). Interestingly, both $Cu_1$-$Co_3O_4$ and $CuO_{NP}$-$Co_3O_4$ exhibit enhanced $CO_2$ evolution (as product of formaldehyde oxidation) compared to pristine $Co_3O_4$, including higher desorption intensities and the appearance of low-temperature desorption features (~230 °C). Pure CuO also shows significant $CO_2$ release, particularly at lower temperatures, indicating that the enhanced formaldehyde reactivity in $CuO_{NP}$-$Co_3O_4$ can be largely attributed to the intrinsic surface reactivity of CuO itself. These observations suggest that both types of Cu species, atomically dispersed and nanoparticulate, contribute similarly to surface adsorption and activation of formaldehyde. Yet, the low-temperature desorption feature appears more pronounced for $Cu_1$-$Co_3O_4$, suggesting that Cu SAs seem to provide a more reactive adsorption environment.

To directly probe electronic interactions between Cu and the lattice-based Co ions, XPS analysis of the Co 2p region was conducted (Figure S8). The deconvolution results in **Figure**

**3c** reveal that the $Co^{2+}/Co^{3+}$ ratio remains largely unchanged in $CuO_{NP}$-$Co_3O_4$ compared to pristine $Co_3O_4$. In contrast, $Cu_1$-$Co_3O_4$ exhibits a significantly higher $Co^{2+}$ fraction, indicating a partial reduction of $Co^{3+}$. This shift suggests electron transfer from $Cu^{2+}$ to the $Co^{3+}$ in the $Co_3O_4$ lattice through interfacial oxygen, facilitated by the intimate atomic-level contact and interfacial Cu–O–Co electronic coupling. These findings support that single-atom Cu sites induce enhanced SMSI compared to nanoparticulate CuO NPs, as schematically illustrated in **Figure 3d**: In $Cu_1$-$Co_3O_4$, each $Cu^{2+}$ is stabilized in direct contact with the $Co_3O_4$ surface, maximizing the interfacial area and enabling efficient and direct charge transfer from $Cu^{2+}$ to the coordinating interfacial oxygen of $Co_3O_4$ lattice. In $CuO_{NP}$-$Co_3O_4$, CuO nanoparticles, only interface-near Cu ions interact with the support, resulting in weaker SMSI and negligible influence on the $Co_3O_4$ surface electronic structure. This agrees with the earlier and more consolidated reduction peak in the $H_2$-TPR profile of $Cu_1$-$Co_3O_4$ (**Figure 3a**), caused by formation of labile oxygen intermediates at the electron-rich Cu–O–Co interface. This ability to modulate the redox behavior of the oxide support at such Cu SA loading (i.e., 1.42 wt%) is particularly relevant for catalysis, as SMSI has been shown to yield exceptional activity in Cu SA systems for oxidation reactions[22] as well as in noble metal SAs under similar interfacial conditions.[48,49]

*2.4. Chemoresistive Formaldehyde Sensing Performance*

To evaluate the practical implications of $Cu_1$-$Co_3O_4$ with its enhanced reactivity for molecular sensing, we investigate its performance toward low-temperature detection of redox-active gas species. Formaldehyde is selected as probing molecule due to its carcinogenic nature[50] and relevance as air pollutant. Therefore, $Cu_1$-$Co_3O_4$ nanoparticles are deposited by spin coating onto an alumina substrate with interdigitated Pt electrodes.

The temperature-dependent sensor response of $Cu_1$-$Co_3O_4$ (red circles), $CuO_{NP}$-$Co_3O_4$ (blue squares) and $Co_3O_4$ (green triangles) to 1 ppm of formaldehyde in dry air is shown in **Figure 4a**. Most importantly, $Cu_1$-$Co_3O_4$ features a high response of 14.5 already at 50 °C, that is more than an order of magnitude higher than $CuO_{NP}$-$Co_3O_4$ (0.5) and $Co_3O_4$ (0.9). This high response at relatively low temperature is particularly striking and aligns well with the enhanced activation of lattice oxygen (**Figure 3a**) and SMSI (**Figure 3b–c**) associated to the Cu SAs. These changes seem to collectively lower the activation energy barrier for formaldehyde oxidation, allowing for an efficient gas–solid interaction at reduced temperature. In contrast, $CuO_{NP}$-$Co_3O_4$ features weaker SMSI, as evidenced by the nearly unchanged Co oxidation state (**Figure 3c**) and TPR profile, similar to $Co_3O_4$. As a result, the optimal sensing temperature of $CuO_{NP}$-$Co_3O_4$ remains higher and the sensor response to formaldehyde is generally lower compared to $Cu_1$-$Co_3O_4$ (**Figure 4a**). Lower operational temperature has practical implications, as it allows for portable or battery-powered gas sensing devices,[51] as needed for instance in medical diagnosis,[52] where minimizing power consumption directly translates to longer device lifetime and broader deployment flexibility.

Remarkably, the $Cu_1$-$Co_3O_4$ features a maximum response of 18.9 at 75 °C. This is consistently higher than other formaldehyde sensors[20,53-62] operated at similarly low temperature (e.g., ≤150 °C), as shown in **Figure 4b**, emphasizing the excellent sensing properties of Cu SAs on $Co_3O_4$. For instance, a $MnO_2$-modified $SnO_2$ sensor exhibited a response of approximately 12 to 1 ppm formaldehyde at 80 °C,[62] and another study demonstrated a porous 3D ZnO structure achieving room-temperature detection of formaldehyde under visible light with a response of 5.6 to 1 ppm.[60] Note that higher response values have been reported only for sensors at higher temperatures (≥250 °C).[16,63]

To challenge our $Cu_1$-$Co_3O_4$ sensor further under more realistic gas environments with relative humidity (RH), we show in **Figure 4c** the sensor resistance change at 75 °C upon consecutive formaldehyde exposure to concentrations between 500 ppb to 5 ppb at 50% RH. The sensor exhibits a well-defined, concentration-dependent response. Even at 5 ppb, that covers the strictest regulatory limits (i.e., 8 ppb in France),[64] a clearly distinguishable response (7.1% with signal-to-noise ratio of ~300) was obtained. Importantly, the resistance baseline is always recovered, indicating fully reversible surface-analyte interaction. We also test the sensor to eleven consecutive cycles of 100 ppb of formaldehyde at 50% RH over 2,500 min (i.e., almost 2 days, Figure S9) and the sensor showed repeatable responses, without any performance degradation.

Selectivity is a critical parameter for indoor air sensing, where formaldehyde must be reliably distinguished from a wide range of volatile organic and inorganic compounds, such as ethanol, acetone, toluene, ammonia, and NO. These compounds are usually found in household, office, or industrial environments at similar or higher concentrations than formaldehyde. Many of these interferents, especially small alcohols or ketones, exhibit overlapping oxidation reactivity on metal oxide surfaces, often resulting in cross-sensitivity[63] and false positives. To address this, we evaluate the absolute sensor response of $Cu_1$-$Co_3O_4$ to several representative indoor interferents at 75 °C and 1 ppm under 50% RH (**Figure 4d**). Among all tested species, the response to formaldehyde is, at least, an order of magnitude and up to 140 times higher. This highlights the intrinsic chemical selectivity of the sensor toward formaldehyde under humid conditions. These results support the potential of chemoresistive sensors based on single-atom isolates with well-defined reactivity for selective analyte detection at low operational temperate. Note that selectivity may be further enhanced with catalytic[65] or sorption-based[66] filtering systems, if needed, to minimize background interference in real-world environments.

**Conclusion**

This work demonstrates that atomic dispersion of non-noble metals can unlock drastically enhanced redox activity in oxide supports, traditionally associated with noble metal systems. By anchoring single Cu atom sites on $Co_3O_4$ we uncover a pronounced shift in lattice oxygen activity and cobalt oxidation states, arising from interfacial Cu–O–Co coupling, compared to analogues of $Co_3O_4$ loaded with CuO nanoparticles. Spectroscopic and temperature-programmed analyses reveal that $Cu_1$-$Co_3O_4$ exhibits enhanced reducibility and oxygen activation at markedly lower temperatures over $CuO_{NP}$-$Co_3O_4$ and $Co_3O_4$, indicating strong electronic perturbation of the support. These redox shifts translate directly into functional performance: $Cu_1$-$Co_3O_4$ sensors show an order of magnitude higher formaldehyde response at temperatures as low as 50 °C, and consistently outperform other state-of-the-art sensors. Importantly, the well-defined reactivity of the Cu SA yields high formaldehyde selectivity over other molecule classes, including alcohols, ketones, aromatic compounds and aldehydes. Together, these findings establish that atomic-scale engineering not only enables fundamental control over support chemistry, but also elevates functional outcomes. Beyond sensors, this redox tuning concept provides a blueprint for designing next-generation oxide-based materials that capitalize on the catalytic influence of earth-abundant single-atom sites.

**Acknowledgement**

This study was financially supported by the Swiss State Secretariat for Education, Research, and Innovation (SERI) under contract number MB22.00041 (ERC-STG-21

"HEALTHSENSE") and by the Swiss National Science Foundation (228416). The authors acknowledge the Diamond Light Source (SP36218) at the B07-C branch for providing synchrotron measurement time, the MAV IV synchrotron at the Balder beamline for providing measurement (20240825), and the Scientific Center for Optical and Electron Microscopy (ScopeM) of ETH Zurich for providing measuring time on their electron microscopes.**Experimental Section**

*Synthesis of pure $Co_3O_4$*: Pure $Co_3O_4$ nanoparticles were synthesized using a flame-spray pyrolysis (FSP) reactor that has been described elsewhere.[67] Briefly, a precursor solution of 0.2 M cobalt(II) 2-ethylhexanoate (65 wt.% in mineral spirits, Sigma Aldrich, Switzerland) dissolved in xylene (isomeric mixture, VWR Chemicals, Switzerland) was introduced through a capillary at a feed rate of 5 mL min$^{-1}$.[35] The solution was atomized by an oxygen dispersion flow of 5 L min$^{-1}$, with a nozzle pressure drop of 1.6 bar, generating a fine spray. This spray underwent oxidation with the help of a flamelet sustained by $CH_4$ (1.25 L min$^{-1}$, Methane 2.5, PanGas, Switzerland) and $O_2$ (3.25 L min$^{-1}$, PanGas, Switzerland), while an additional $O_2$ sheath flow (5 L min$^{-1}$) provided flame stabilization. The resulting nanoparticles were collected on a water-cooled glass fiber filter (257 mm diameter, GF6, Hahnemühle Fineart, Germany) positioned 57 cm above the burner, with vacuum assistance (Seco SV 1025 C, Busch, Switzerland). The collected powder was retrieved by scraping the filter with a spatula, followed by sieving through a 250 μm stainless steel mesh to eliminate residual filter fibers. Obtained powders were annealed at 500 °C and air atmosphere in an oven (CWF 1300, Carbolite Gero, Germany) before characterizations and gas sensor testing.

*Synthesis of $Cu_1$-$Co_3O_4$*: Wet impregnation was used to stabilize Cu single-atoms onto the pure, unannealed $Co_3O_4$ nanoparticles. 100 mg of pure FSP-made and non-annealed $Co_3O_4$ nanoparticles and 36.6 mg of $Cu(NO_3)_2·2.5H_2O$ (ACS reagent, 98%, Sigma Aldrich, Switzerland) were dispersed in 10 mL deionized water. The dispersion of the nanoparticles in aqueous $Cu(NO_3)_2$ solution was vigorously stirred with magnetic stirrer for 6 hours before being washed and centrifugated several times with DI water and ethanol. Powders collected after centrifugation were dried in an oven overnight at 70 °C, then annealed in a furnace at 500 °C for 5 hours under static air.

*Synthesis of $CuO_{NP}$-$Co_3O_4$*: First, pure CuO nanoparticles were prepared using the same protocol with the FSP reactor, except that the precursor solution is composed of 0.2 M of Deca

Copper 8 (Borchers, Germany) in xylene. Also the feeding rate of the precursor solution was reduced to 2 mL min$^{-1}$. To prepare CuO$_{NP}$-Co$_3$O$_4$ nanoparticles, the same protocol was used as for the wet impregnation in the synthesis of Cu$_1$-Co$_3$O$_4$, by replacing the Cu(NO$_3$)$_2$ precursor with CuO nanoparticles with the same atomic loading of Cu.

*Chemoresistive gas sensor preparation and measurement*: 6 mg of annealed pure Co$_3$O$_4$, Cu$_1$-Co$_3$O$_4$ and CuO$_{NP}$-Co$_3$O$_4$ were each dispersed in 120 μL of ethanol by ultrasonication. Dispersed solutions were each completely dropped onto alumina substrates with interdigitated electrodes and a back heater (electrode type #103, Electronic Design Center, Case Western University, USA), and spin-coated at 250 RPM for 30 minutes. The resulting films were dried in an oven at 70 °C for 30 minutes.

The sensors were affixed to Macor holders and positioned within a Teflon chamber, as detailed in previous studies.[68] The operating temperature was regulated by applying a constant voltage to the Pt heater embedded in the alumina sensing substrate. The chamber was integrated into a gas mixing system[69] via inert Teflon tubing. Hydrocarbon-free synthetic air (PanGas, C$_x$H$_y$ and NO$_x$ < 100 ppb, Switzerland) was employed as the carrier gas. Certified gas mixtures (PanGas, Switzerland) were blended using calibrated mass flow controllers (Bronkhorst, Netherlands) to achieve the target composition. The gas standards included: acetone (14.9 ppm, PanGas, Switzerland), toluene (9.6 ppm, PanGas, Switzerland), ethanol (15.0 ppm, PanGas, Switzerland), ammonia (10.3 ppm, PanGas, Switzerland), methane (10.2 ppm, all in synthetic air, Pangas, Switzerland), NO (10.2 ppm, PanGas, Switzerland), acetaldehyde (17.1 ppm, PanGas, Switzerland), and formaldehyde (17.0 ppm all in N$_2$, PanGas, Switzerland). To generate humidified air, dry synthetic air was passed through a bubbler containing deionized water at room temperature (ca. 25 °C), and the resulting moisture-laden stream was blended with the analyte-containing gas flow to achieve the target relative humidity. The total gas flow rate was maintained at 300 mL min$^{-1}$. The dynamic ohmic resistance of the sensing film deposited on the electrodes was monitored using a digital multimeter (Keithley 2700, Keithley Instruments, USA). The chemoresistive sensor response S was defined by the equation below:

$$S = \frac{R_g}{R_a} - 1$$

Where $R_g$ and $R_a$ are the resistances of the sensing layer under gas exposure and in air, respectively. The signal-to-noise ratio (SNR) of the $Cu_1$-$Co_3O_4$ sensor at 5 ppb formaldehyde was evaluated as:

$$SNR = \frac{\Delta R}{Noise}$$

Where ΔR is the change in resistance toward exposure to 5 ppb of formaldehyde and the noise is the standard deviation of the baseline readout over 3 min prior to exposure.

*Materials characterization*: XRD patterns of powders were acquired with a Bruker D2 Phaser (USA) operated at 30 kV and 10 mA (Cu $K_\alpha$ radiation, λ = 1.5406 Å), with scanning step size of 0.01° and a scanning time of 2.2 seconds per step.

X-ray absorption spectroscopy measurements were taken at Balder beamline of MAX IV (Lund, Sweden) with transmission and fluorescence XANES and EXAFS in the energy range from 8–10 keV (double crystal monochromator Si(111)) at room temperature. Samples were prepared as pressed pellets (13 mm diameter) by homogenizing the catalyst with boron nitride (1:5 mass ratio) to minimize self-absorption effects. Analyses of the near edge and extended range were carried out using Athena software, and the fitting of each sample was carried out by Artemis software. EXAFS spectra were fitted in a Fourier-transform range of 3–11 Å$^{-1}$ with a Hanning window applied between 1 Å and 3 Å. NEXAFS measurements were carried out under vacuum at the B07 beamline of Diamond Light Source (UK). Reference samples (CuO, 99.99% trace metal basis, and $Cu_2O$, ≥99.99% trace metals basis, anhydrous) were purchased from Sigma Aldrich, Switzerland. XPS was carried out using Al Kα radiation (Sigma Probe, Thermo VG Scientific, USA). Surface morphology and cross-section SEM images of the nanoparticles were carried out by field-emission SEM under 2.0 kV and 0.10 nA (Thermofisher Scientific, Magellan 400, USA).

Particle characterization was carried out using transmission electron microscopy (TEM) and scanning transmission electron microscopy (STEM) on a JEOL JEM-F200 equipped with a cold field emitter as electron source, operated at 200 kV (JEOL, Japan). Samples for imaging were prepared by dispersing the particles in ethanol and depositing them onto perforated carbon films supported on molybdenum grids. For compositional analysis, energy-dispersive X-ray spectroscopy (EDXS) was performed in STEM mode utilizing four silicon drift detectors (SDDs) from JEOL. High-resolution TEM (HR-TEM) imaging was conducted on a Grand

ARM300F microscope (JEOL, Japan) operating at 300 kV with a cold field emission source, where aberration correction for both the image-forming and probe-forming lenses allowed sub-angstrom resolution in TEM and STEM modes. Quantitative analysis of Cu loading was performed by inductively coupled plasma mass spectrometry (ICP–MS, Agilent 7900, Agilent Technologies, USA). Samples (~5 mg) were digested in a mixture of $HNO_3$ and HCl using microwave-assisted acid digestion (overnight) prior to measurement.

Temperature-programmed experiments were performed with an Autochem III chemisorption analyzer (Micromeritics, USA), equipped with a TCD detector and connected to a quadrupole mass spectrometer (Omnistar, Pfeiffer, Germany). For $H_2$-TPR, approx. 40 mg of powders were pretreated under Ar at 200 °C. The reduction was performed under 30 mL/min of 5 vol% $H_2$ in Ar, between 40 - 600 °C at a rate of 5 °C/min. A moisture trap ensured that no humidity reached the TCD, as also confirmed by mass spectrometry. For formaldehyde-TPD, approx. 60 mg of powders were pretreated in He at 200 °C. Formaldehyde (10 ppm in $N_2$, Pangas, Switzerland) was supplied at a rate of 10 mL/min at 30 °C for 120 minutes. Thereafter, desorption was carried out under 10 mL/min He between 30 - 600 °C at a rate of 5 °C/min, and analyzed by mass spectrometry at m/z ratios of 2, 18, 28, 29, 30, 44.

*Statistical Methods*: The average crystallite sizes of the $Co_3O_4$-based samples were estimated using the Scherrer equation, applied to the 311, 511, and 440 diffraction planes identified in the XRD patterns. The full width at half maximum (FWHM) values were extracted after background subtraction and instrumental broadening correction. The crystallite sizes calculated from the three reflections are indicated as average and standard deviation to obtain a representative value for each sample.

XPS spectra were processed and fitted using the Thermo Scientific Avantage software. Background subtraction was performed using the Shirley method, which is the standard approach in Avantage for core-level spectra and accounts for inelastic background contributions. In cases where the Shirley background was not applicable due to spectral shape, a linear background was used as implemented in Avantage. Peak fitting was carried out using a non-linear least squares algorithm with a combination of Gaussian–Lorentzian (Voigt) peak shapes, allowing peak positions, widths, and intensities to vary within physically reasonable limits. All processing steps, including background subtraction and peak fitting, were recorded in the Avantage audit trail for reproducibility. Quantification was based on the integrated peak areas after background subtraction, applying the instrument transmission function and appropriate sensitivity factors as provided by the Avantage library.

EXAFS fitting was performed on the Cu K-edge spectra using the Artemis software package. The theoretical paths were generated from the FEFF6 code based on a $Co_3O_4$ lattice model with Cu substitution. For $Cu_1$-$Co_3O_4$, fitting was carried out using a two-shell model: a Cu–O shell with a coordination number of 3.5 and a Cu–Co shell at ~2.89 Å with CN ≈ 1. No Cu–Cu scattering was included, consistent with the absence of Cu clustering. Reference compounds (CuO, $Cu_2O$, Cu foil) were also fitted for comparison. Good agreement between experimental and fitted spectra is shown in Figure S6.

2981-2997 (2023). https://doi.org/10.1021/acscatal.2c05992

14  Vorobyeva, E. *et al.* Atom-by-Atom Resolution of Structure–Function Relations over Low-Nuclearity Metal Catalysts. *Angewandte Chemie International Edition* **58**, 8724-8729 (2019). https://doi.org/https://doi.org/10.1002/anie.201902136

15  Bulemo, P. M. *et al.* Selectivity in Chemiresistive Gas Sensors: Strategies and Challenges. *Chemical Reviews* **125**, 4111-4183 (2025). https://doi.org/10.1021/acs.chemrev.4c00592

16  Shin, H. *et al.* Single-Atom Pt Stabilized on One-Dimensional Nanostructure Support via Carbon Nitride/SnO2 Heterojunction Trapping. *ACS Nano* **14**, 11394-11405 (2020). https://doi.org/10.1021/acsnano.0c03687

17  Shin, H. *et al.* Sacrificial Template-Assisted Synthesis of Inorganic Nanosheets with High-Loading Single-Atom Catalysts: A General Approach. *Advanced Functional Materials* **32**, 2110485 (2022). https://doi.org/https://doi.org/10.1002/adfm.202110485

18  Koga, K. Electronic and Catalytic Effects of Single-Atom Pd Additives on the Hydrogen Sensing Properties of Co3O4 Nanoparticle Films. *ACS Applied Materials & Interfaces* **12**, 20806-20823 (2020). https://doi.org/10.1021/acsami.9b23290

19  Ye, X.-L. *et al.* Boosting Room Temperature Sensing Performances by Atomically Dispersed Pd Stabilized via Surface Coordination. *ACS Sensors* **6**, 1103-1110 (2021). https://doi.org/10.1021/acssensors.0c02369

20  Gu, F., Di, M., Han, D., Hong, S. & Wang, Z. Atomically Dispersed Au on In2O3 Nanosheets for Highly Sensitive and Selective Detection of Formaldehyde. *ACS Sensors* **5**, 2611-2619 (2020). https://doi.org/10.1021/acssensors.0c01074

21  Xue, Z. *et al.* One-Dimensional Segregated Single Au Sites on Step-Rich ZnO Ladder for Ultrasensitive NO2 Sensors. *Chem* **6**, 3364-3373 (2020). https://doi.org/https://doi.org/10.1016/j.chempr.2020.09.026

22  Mosrati, J. *et al.* Tiny Species with Big Impact: High Activity of Cu Single Atoms on CeO2–TiO2 Deciphered by Operando Spectroscopy. *ACS Catalysis* **11**, 10933-10949 (2021). https://doi.org/10.1021/acscatal.1c02349

23  Abdel-Mageed, A. M. *et al.* Unveiling the CO Oxidation Mechanism over a Molecularly Defined Copper Single-Atom Catalyst Supported on a Metal–Organic Framework. *Angewandte Chemie International Edition* **62**, e202301920 (2023). https://doi.org/https://doi.org/10.1002/anie.202301920

24  Li, F. *et al.* Boosting oxygen reduction catalysis with abundant copper single atom active sites. *Energy & Environmental Science* **11**, 2263-2269 (2018). https://doi.org/10.1039/C8EE01169A

25  Qu, Y. *et al.* Direct transformation of bulk copper into copper single sites via emitting and trapping of atoms. *Nature Catalysis* **1**, 781-786 (2018). https://doi.org/10.1038/s41929-018-0146-x18

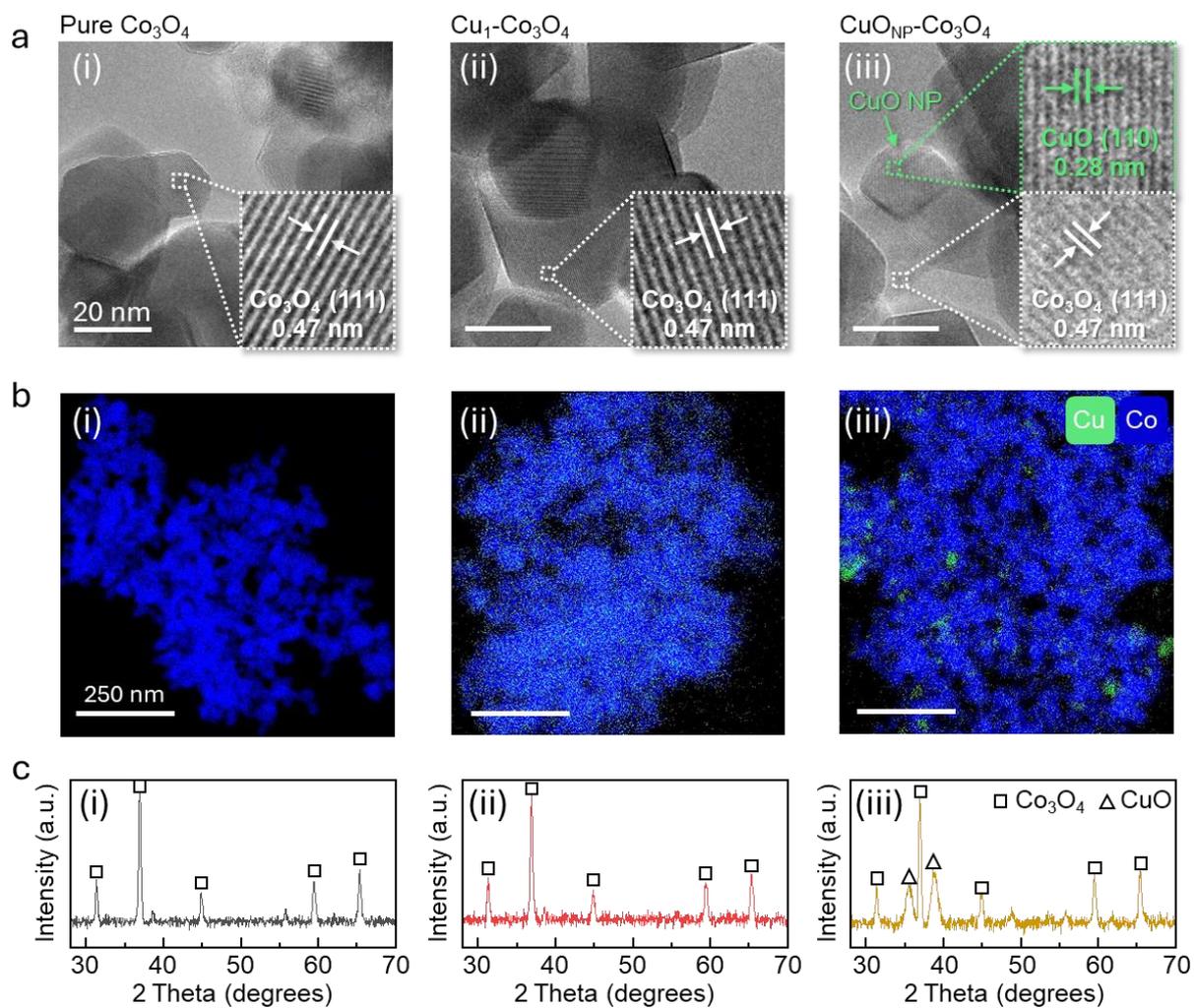

**Figure 1.** a) High-resolution TEM images with magnifications and labelled lattice fringes in the insets of (i) pristine $Co_3O_4$, (ii) $Cu_1$-$Co_3O_4$, and (iii) $CuO_{NP}$-$Co_3O_4$. b) STEM-EDS elemental mapping images indicating Co (blue) and Cu (green). c) Powder XRD patterns.

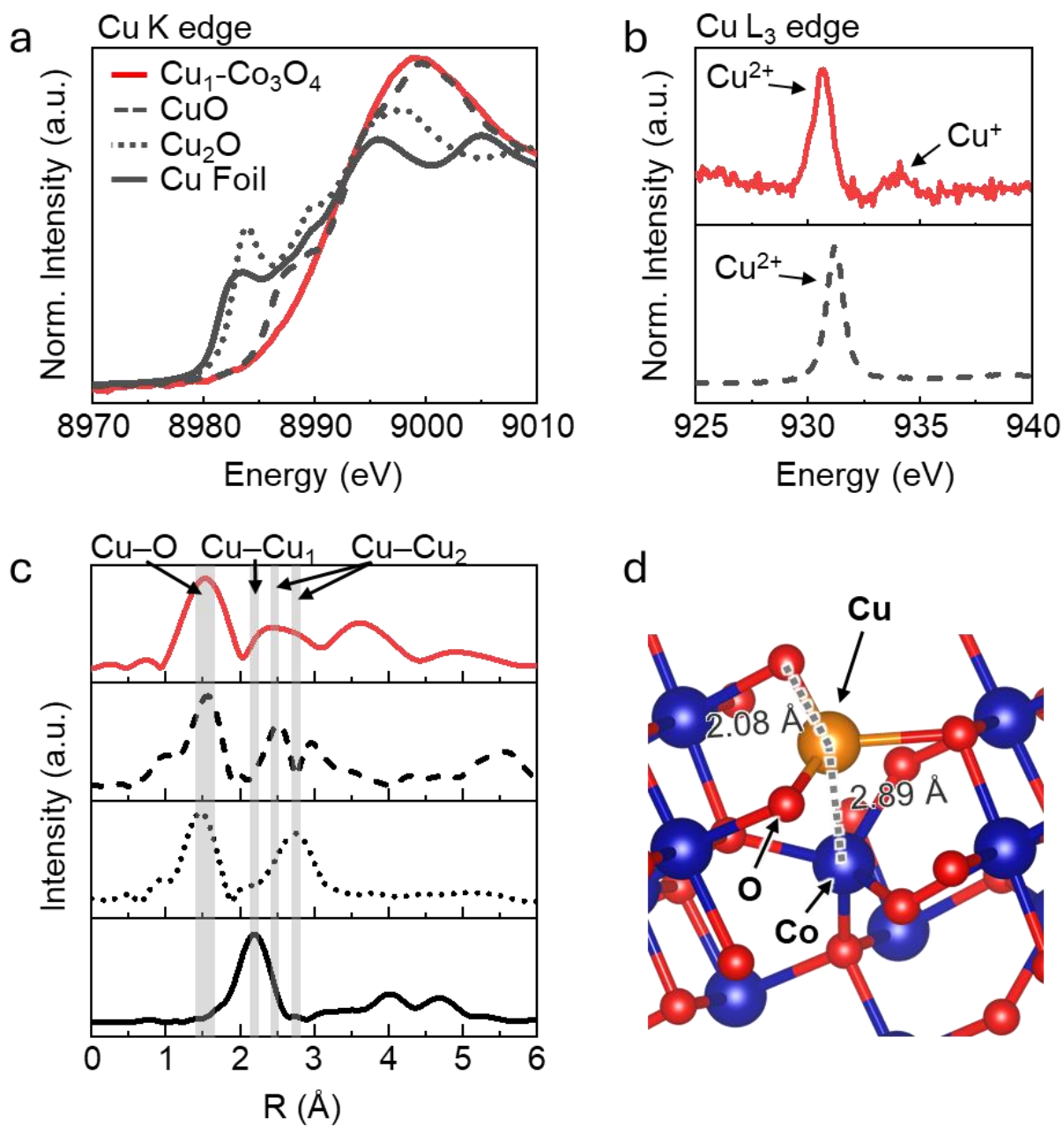

**Figure 2.** a) XANES spectra of $Cu_1$-$Co_3O_4$ and reference powder/foil samples. b) NEXAFS spectrum of $Cu_1$-$Co_3O_4$ and reference CuO powder at the Cu $L_3$-edge under ultra-high vacuum. c) Fourier-transformed Cu K-edge EXAFS spectra of $Cu_1$-$Co_3O_4$ and various reference samples. Displayed by gray-shaded areas are the positions of Cu–O,[45] first shell Cu–Cu and second shell Cu–Cu.[26] d) Atomic model of Cu SA (orange) anchored on $Co_3O_4$ surface, showing average Cu-O and Cu-Co bond distances of 2.08 Å and 2.89 Å, respectively, as indicated by dotted lines.

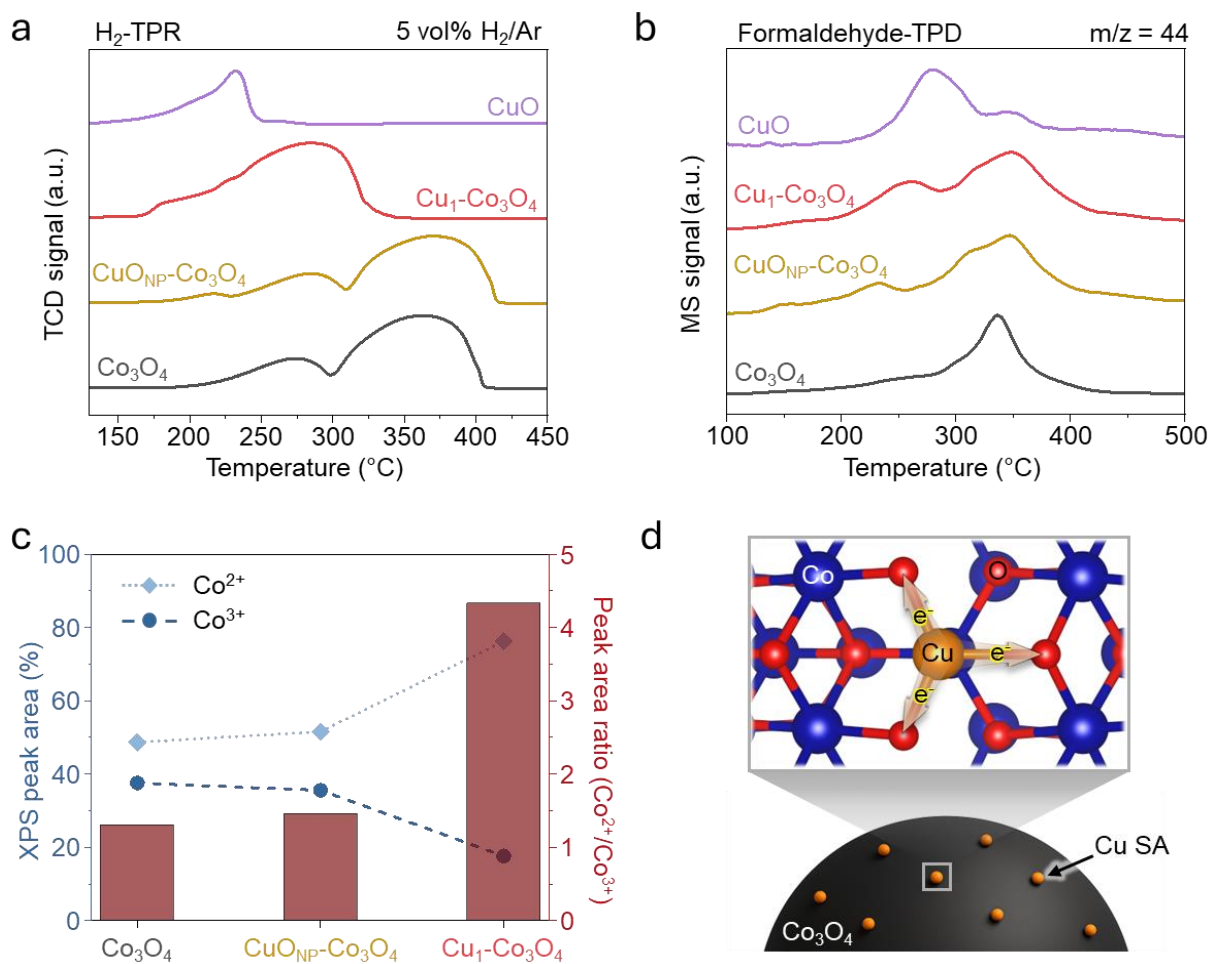

**Figure 3.** a) H$_2$-TPR profiles of pure Co$_3$O$_4$, CuO$_{NP}$-Co$_3$O$_4$, Cu$_1$-Co$_3$O$_4$ and CuO under 5 vol% H$_2$/Ar. b) Formaldehyde-TPD profiles (m/z = 44, CO$_2$) of pure Co$_3$O$_4$, CuO$_{NP}$-Co$_3$O$_4$ Cu$_1$-Co$_3$O$_4$ and CuO under 10 ppm formaldehyde in dry air. c) Peak areas of Co$^{3+}$ and Co$^{2+}$ (symbols, left ordinate), and their ratios (bars, right ordinate) in pure Co$_3$O$_4$, CuO$_{NP}$-Co$_3$O$_4$ and Cu$_1$-Co$_3$O$_4$ determined from XPS analysis (Figure S8). d) Schematic illustration showing a single Cu atom (orange) coordinated to surface oxygen atoms (red). Such Cu sites are distributed on the surface of a Co$_3$O$_4$ nanoparticle.

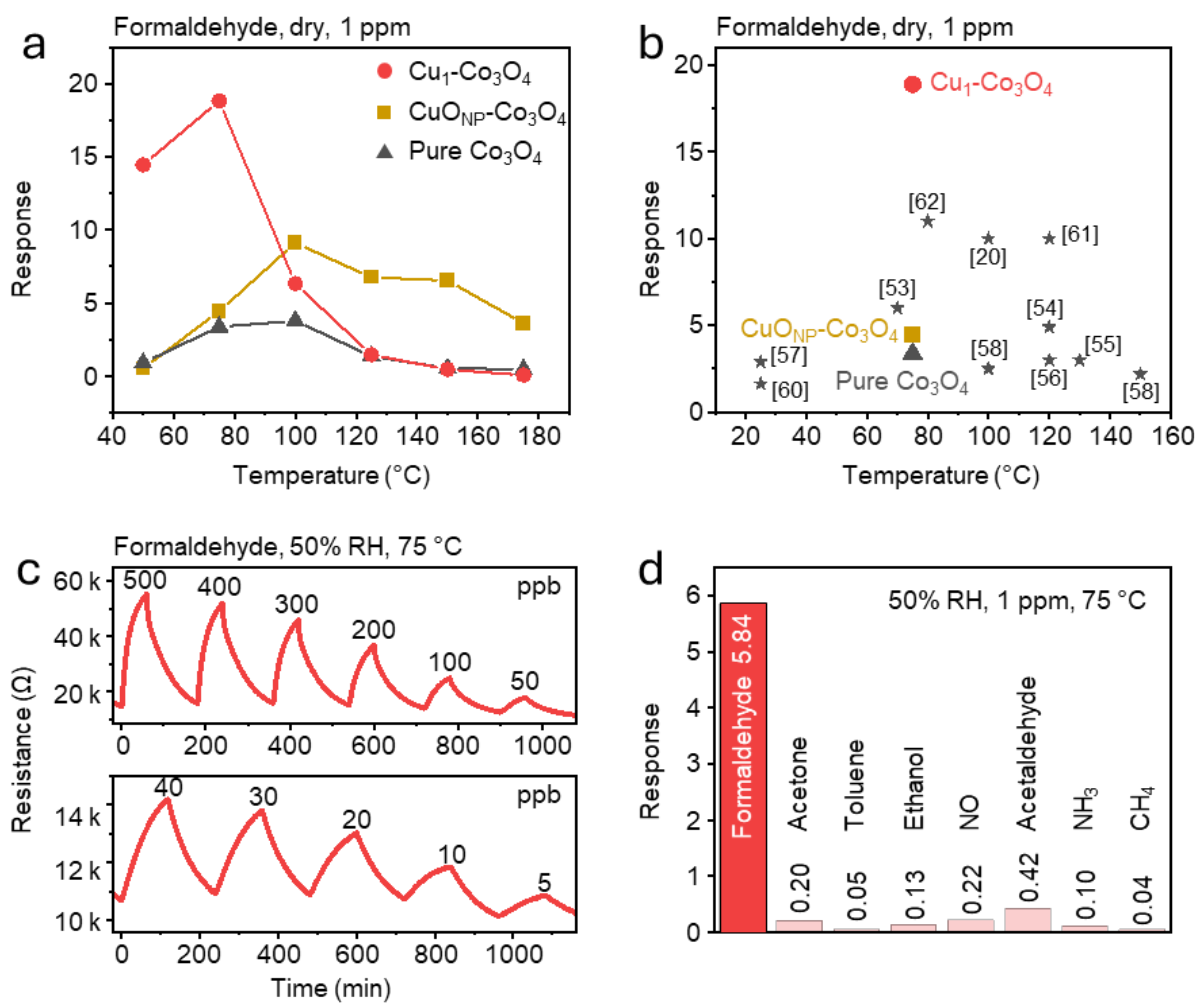

**Figure 4.** a) Temperature-dependent response of $Cu_1$-$Co_3O_4$ and reference samples toward 1 ppm formaldehyde under dry air condition between 50–170 °C. b) Comparison of the formaldehyde sensing responses (at 1 ppm) of $Cu_1$–$Co_3O_4$ and representative formaldehyde sensors reported in the literature that operate below 150 °C.[20,53-62] c) Dynamic resistance profiles of $Cu_1$-$Co_3O_4$ sensor upon exposure to varying concentrations of formaldehyde (5 to 500 ppb) at 50% RH and 75 °C. d) $Cu_1$-$Co_3O_4$ sensor response to formaldehyde and various common indoor air interferents under 50% RH and 1 ppm.

**Supplementary Information**

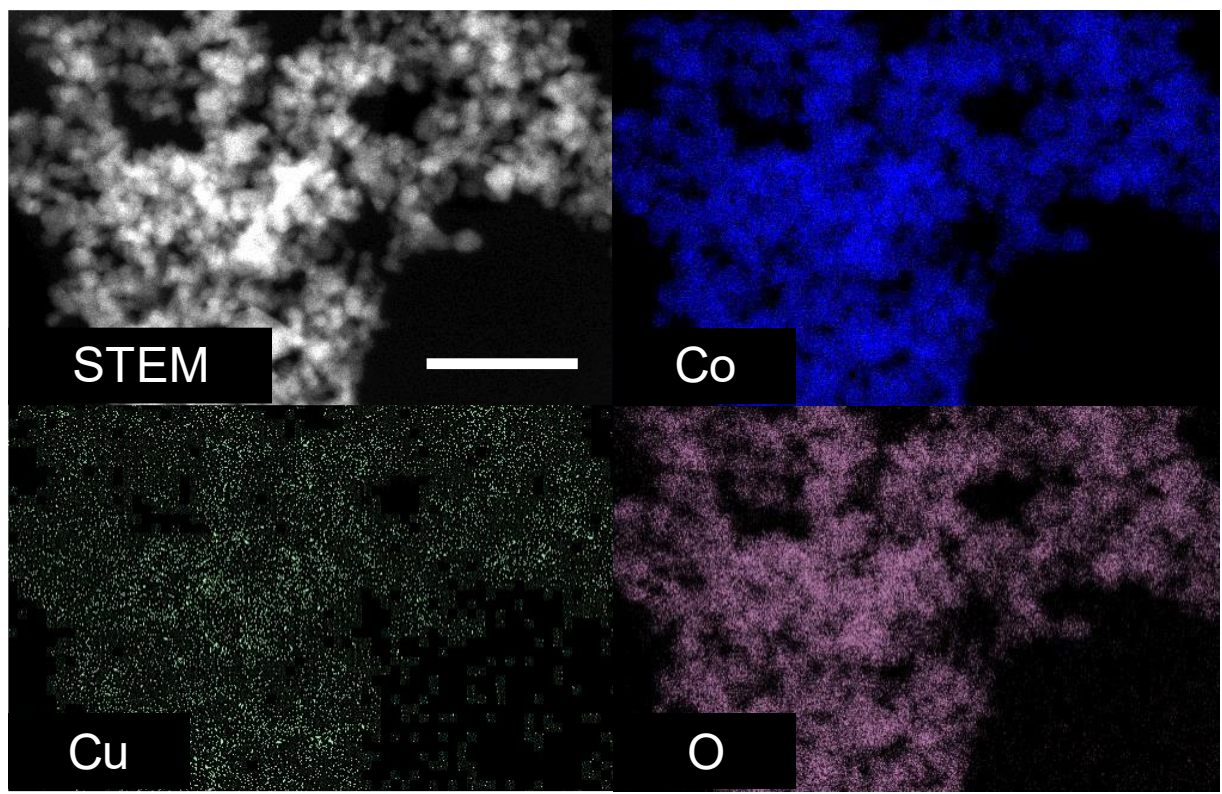

**Figure S1:** STEM image of $Cu_1$-$Co_3O_4$ and corresponding EDS mapping of Co, Cu, and O. Scale bar: 250 nm.

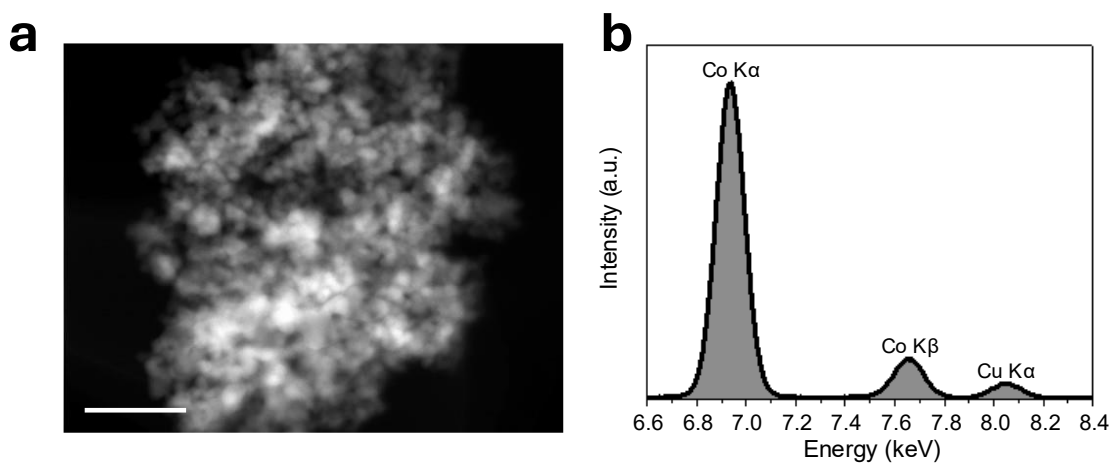

**Figure S2:** a) Scanning transmission electron microscopy (STEM) image of $Cu_1$-$Co_3O_4$ nanoparticles. Scale bar: 200 nm. b) Energy-dispersive X-ray spectroscopy (EDS) spectrum collected from (a), confirming the elemental composition.

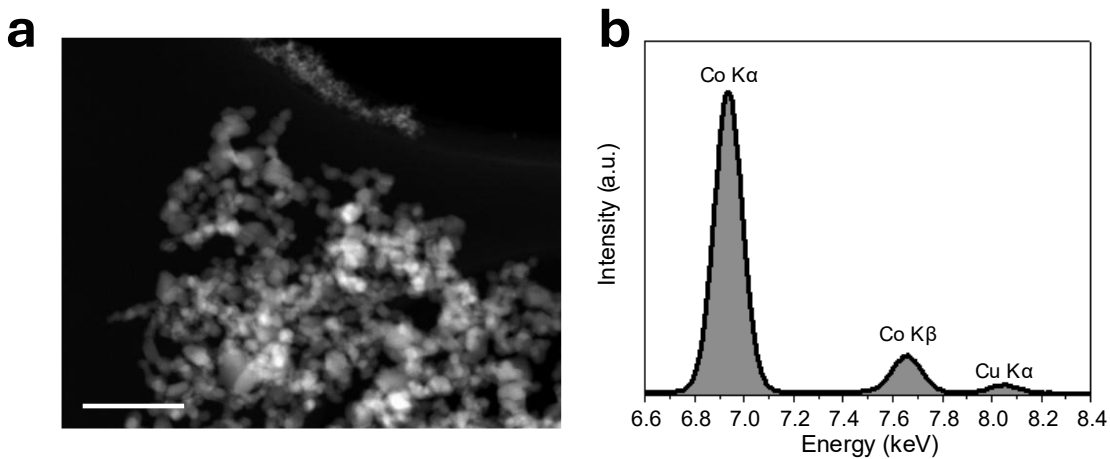

**Figure S3:** a) Scanning transmission electron microscopy (STEM) image of $Cu_{NP}$-$Co_3O_4$ nanoparticles. Scale bar: 200 nm. b) Energy-dispersive X-ray spectroscopy (EDS) spectrum collected from (a), confirming the elemental composition.

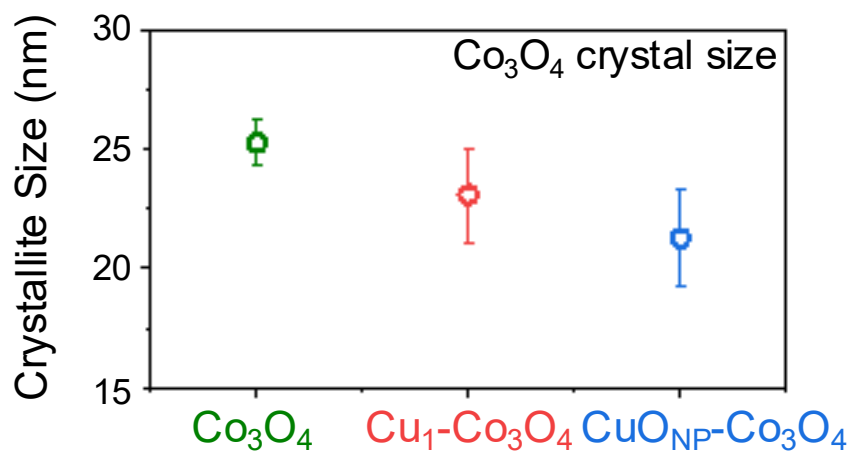

**Figure S4:** $Co_3O_4$ Crystallite sizes of pure $Co_3O_4$, $Cu_1$-$Co_3O_4$, and $CuO_{NP}$-$Co_3O_4$. Symbols indicate average and error bars standard deviation of the crystal size estimated by the Scherrer equation from the (311), (511), and (440) planes.

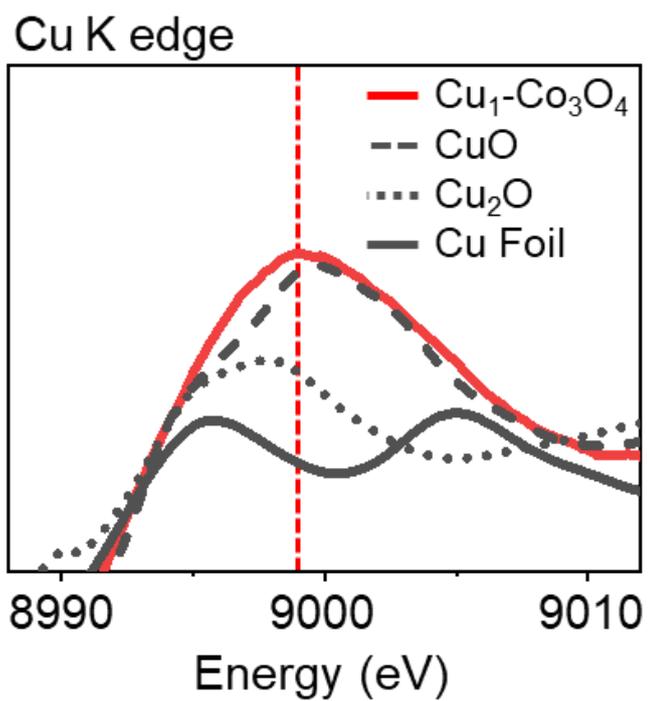

**Figure S5:** Magnified XANES spectra of $Cu_1$-$Co_3O_4$ and reference samples at Cu K edge. The red dashed vertical line indicates the position of the absorption peak of Cu species in $Cu_1$-$Co_3O_4$.

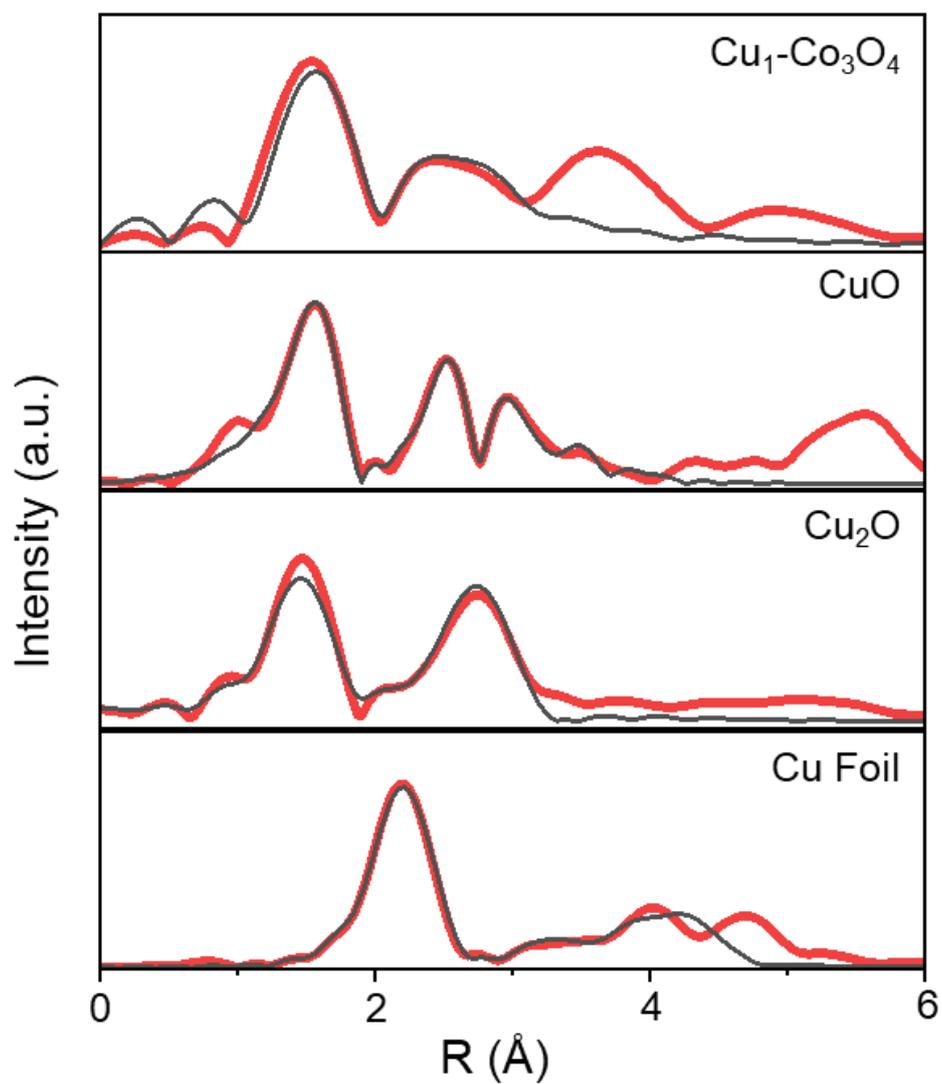

**Figure S6:** EXAFS curve fittings of $Cu_1$-$Co_3O_4$ and reference samples. Red lines are the experimental data, and black lines are the fit.

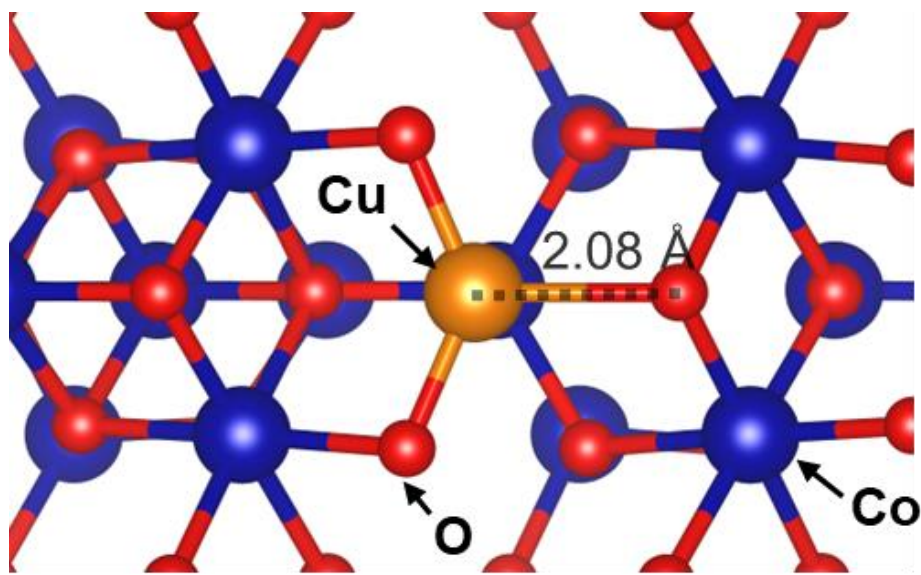

**Figure S7:** Alternative view of the atomic model of Cu SA anchored on Co$_3$O$_4$ surface.

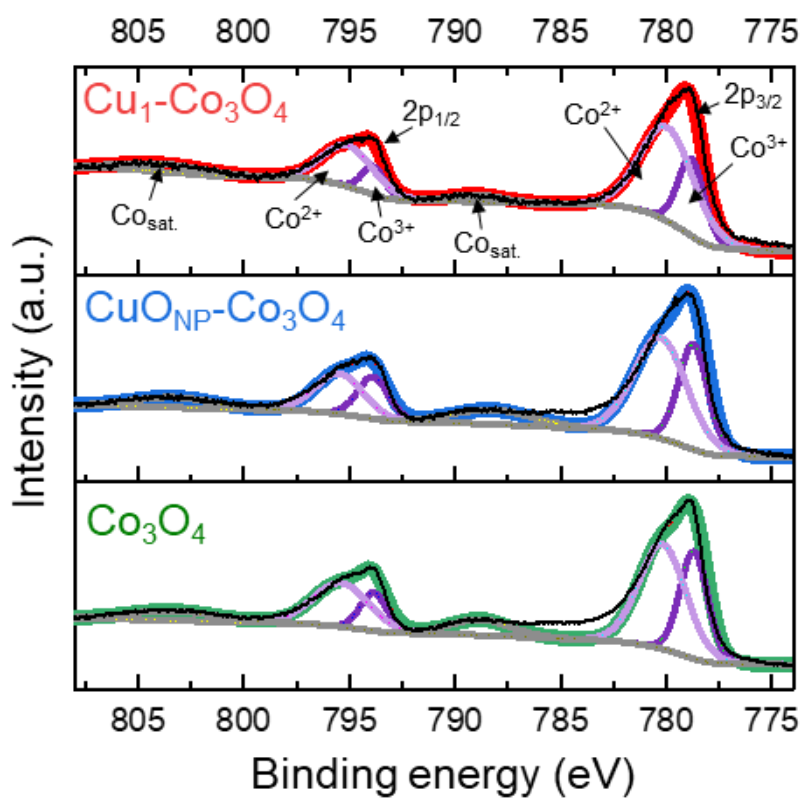

**Figure S8:** XPS spectra of Cu$_1$-Co$_3$O$_4$, CuO$_{NP}$-Co$_3$O$_4$, and Co$_3$O$_4$ in the vicinity of Co 2p. Black line is the experimental data, and the red, blue and green lines are the fit data for Cu$_1$-Co$_3$O$_4$, CuO$_{NP}$-Co$_3$O$_4$ and pure Co$_3$O$_4$, respectively. Light purple and dark purple are the Co$^{2+}$ and Co$^{3+}$ peak fits, respectively.

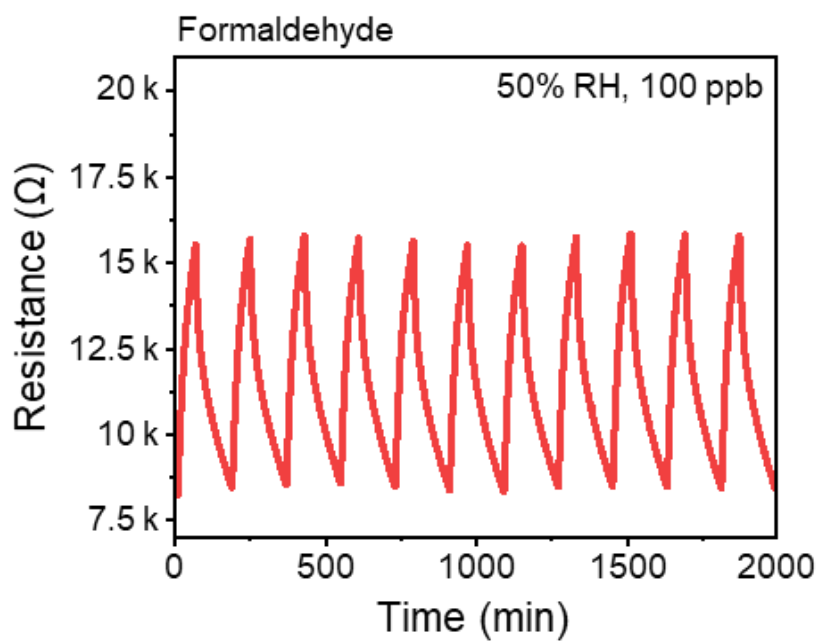

**Figure S9:** Stability of $Cu_1$-$Co_3O_4$ under 11 repeated exposure and recovery to 100 ppb formaldehyde.

**Table S1:** EXAFS fitting table. CN: coordination number, R: bond distance, $\sigma^2$: Debye-Weller factor. Cu–M is either Cu–Cu or Cu–Co.

| Sample | Shell | CN | R (Å) | $\sigma^2$ (Å$^2$) | R-factor |
|---|---|---|---|---|---|
| $Cu_1$-$Co_3O_4$ | Cu–O | 3.5 | 2.08±0.051 | 0.037±0.029 | 0.009 |
|  | Cu–M | 1 | 2.89±0.023 | 0.007±0.016 |  |
| CuO | Cu–O | 4 | 1.95±0.004 | 0.005±0.001 | 0.006 |
|  | Cu–Cu | 4 | 2.94±0.008 | 0.005±0.001 |  |
| $Cu_2O$ | Cu–O | 2 | 1.85±0.013 | 0.005±0.003 | 0.019 |
|  | Cu–Cu | 12 | 3.02±0.016 | 0.020±0.002 |  |
| Cu Foil | Cu–Cu | 12 | 2.54±0.001 | 0.009±0.001 | 0.006 |